\documentclass[aps,showpacs,twocolumn]{revtex4}
\usepackage{epsfig}
\usepackage{amsmath}

\begin{document}
\title{Heavy pentaquark states and a novel color structure}

\author{Chengrong Deng$^{1}{\footnote{crdeng@cqjtu.edu.cn}}$,
        Jialun Ping$^2{\footnote{jlping@njnu.edu.cn, corresponding author}}$,
        Hongxia Huang$^{2}{\footnote{hxhuang@njnu.edu.cn}}$,
        and Fan Wang$^3{\footnote{fgwang@chenwang.nju.edu.cn}}$}
\affiliation{$^1$Department of Physics, Chongqing Jiaotong University, Chongqing 400074, China}
\affiliation{$^2$Department of Physics, Nanjing Normal University, Nanjing 210097, China}
\affiliation{$^3$Department of Physics, Nanjing University, Nanjing 210093, China}

\begin{abstract}

Encouraged by the observation of the pentaquark states $P^+_c(4380)$ and $P^+_c(4450)$, we propose a novel color flux-tube structure,
pentagonal state, for pentaquark states within the framework of color flux-tube mode involving a five-body confinement potential.
Numerical results on the heavy pentaquark states indicate that the states with three color flux-tube structures, diquark, octet
and pentagonal structures, have the close masses, which can therefore be called QCD isomers analogous to isomers in Chemistry.
The pentagonal structure has lowest energy. The state $P^+_c(4380)$ can be described as the compact pentaquark state $uudc\bar{c}$
with the pentagonal structure and $J^P=\frac{3}{2}^-$ in the color flux-tube model.  The state $P^+_c(4450)$ can not be accommodated
into the color flux-tube model. The heavy pentaquark states $uudc\bar{b}$, $uudb\bar{c}$ and $uudb\bar{b}$ are predicted in the color
flux-tube model. The five-body confinement potential basing on the color flux-tube picture as a collective degree of freedom is a
dynamical mechanism in the formation of the compact heavy pentaquark states.

\end{abstract}


\maketitle

\section{Introduction}
In the constituent quark models, conventional hadrons, baryons and mesons, are assumed to be composed of three valence quarks $qqq$ and
a valence quark $q$ and a valence antiquark $\bar{q}$, respectively. Quantum chromodynamics (QCD) does not deny the existence of exotic
hadrons besides the $q\bar{q}$-meson and $qqq$-baryon paradigm. Searching for exotic hadrons has been one of the most significant research
topics of hadronic physics since the pioneer work by Gell-Mann~\cite{pioneer}, in which mesons and baryons can also be, respectively,
tetraquark and pentaquark states if the excitation of a sea quark pair $q\bar{q}$ is taken into account. Exotic hadrons, if they really
exist, may contain more information about the low-energy QCD than that of conventional hadrons. In recent years, a number of experiments
have been witnessing the proliferation of the member of exotic hadron family. The charged tetraquark states $Z_b$~\cite{zb} and $Z_c$~\cite{zc}, dibaryon resonance state $d^*$~\cite{d*}, tetraquark state $X(5568)$~\cite{x5568} and charmonium-pentaquark states $P^+_c(4380)$ and $P^+_c(4450)$~\cite{pc} have been giving us a stimulating glance into the abundant multiquark hadronic world and providing an excellent
opportunity to explore the fundamental freedom playing an essential role in the multiquark hadron states and hadron-hadron interaction.

The hidden charmed states $P^+_c(4380)$ and $P^+_c(4450)$ were recently reported by LHCb Collaboration in the $J/\psi p$ invariant mass
spectrum in the $\Lambda_b^0\rightarrow J/\psi K^-p$ process~\cite{pc}. Their masses and decay widths from a fit using Breit-Wigner
amplitudes are
\begin{eqnarray}
M_{4380}&=&4380^{+8+29}_{-8-29}~\mbox{MeV},~\Gamma_{4380}=205^{+18+86}_{-18-86}~\mbox{MeV},\nonumber\\
M_{4450}&=&4449.8^{+1.7+2.5}_{-1.7-2.5}~\mbox{MeV},~\Gamma_{4450}=39^{+5+19}_{-5-19}~\mbox{MeV}\nonumber
\end{eqnarray}
The $J/\psi p$ decay modes of the two $P^+_c$ states suggest that, regardless of their internal dynamics, they must have minimum intrinsic
quark content $uudc\bar{c}$ with isospin $I=\frac{1}{2}$. However, their total angular momentum and parity $J^P$ cannot be completely
determined up till now, which may be $(\frac{3}{2}^-,\frac{5}{2}^+)$, $(\frac{3}{2}^+,\frac{5}{2}^-)$ or $(\frac{5}{2}^+,\frac{3}{2}^-)$.
A large amount of interpretations in different theoretical frameworks have therefore been proposed to reveal the underlying structures
of these two pentaquark states so far, such as meson-baryon molecule states~\cite{molecule}, diquark-diquark-antiquark states~\cite{diquark},
compact and loose diquark-triquark states~\cite{diquark-triquark}, kinematic effects~\cite{kinematiceffects}, nucleon-$\psi(2S)$ bound state~\cite{npsi}, proton-$\chi_{c1}$ state~\cite{p-chi}, etc. What is the eventually true physical picture of these two pentaquark states?
Further experimental and theoretical work are therefore needed to clear the current complicated situation. In addition, the large mass of
the pentaquark states $P^+_c(4380)$ and $P^+_c(4450)$ mainly comes from the large masses of the heavy charm quark and antiquark $c\bar{c}$. Consequently, a natural question is that what could be the analogous heavy pentaquark states, such as $uudc\bar{b}$, $uudb\bar{c}$ and
$uudb\bar{b}$.

QCD has been widely accepted as the fundamental theory to describe the interactions among quarks and gluons and the structure of hadrons,
in which color confinement is a long distance behavior whose understanding continues to be a challenge in the theoretical physics. It is
well known that color flux-tube-like structures emerge by analyzing the chromo-electric fields between static quarks in lattice numerical simulations~\cite{lns}. Such color flux-tube structures naturally lead to a linear confinement potential between static color charges and
to a direct numerical evidence of color confinement~\cite{linear}. A color flux-tube starts from each quark and ends at an antiquark or a
Y-shaped junction, where three flux tubes are either annihilated or created~\cite{junction}. The color flux-tube structures for mesons and
baryons are seem to be unique and simple. A quark and an antiquark in mesons are connected through a color flux tube. Three quarks in baryons
are connected by a Y-shaped color flux-tube into a color singlet. In general, a multiquark state with $N+1$ particles can be generated by
replacing a quark or an antiquark in an $N$-particle state by a Y-shaped junction and two antiquarks or two quarks. In this way, any multiquark
state must possesses a large number of different topological structures of internal color flux-tube configurations.

It is a well-known fact that the nuclear force in the QCD world and the molecule force in the quantum electrodynamics (QED)
world are very similar except for the length and energy scale difference. Furthermore, the color flux tubes in a hadrons
should also be very analogous to the chemical bond in a molecule. Like the organic world full of variety because of the
chemical bonds, i.e. isomers, the multiquark hadron world may be equally or even more diverse due to the color flux-tube
structure, which can be similarly called QCD isomeric compounds here. Theoretically, QCD is more complicated than QED
so that it is natural to expect that the structures of QCD matters are abundant, even more various than that of QED matters.

In the previous work, we advanced possible color flux-tube structures, so-called QCD quark cyclobutadiene and QCD benzene, for
tetra-quark and six-quark states respectively within the framework of color flux-tube model basing on the lattice QCD (LQCD) picture
and traditional quark models~\cite{cyclobutadiene,ping}. In the paper, we propose and study a novel color flux-tube structure, called
pentagonal state, for the heavy pentaquark states to attempt to enrich the knowledge of inner structures of multiquark states. In
addition, the heavy pentaquark states are also systematically investigated in the color flux-tube model, which may be useful for
exploring exotic baryons in future experiments.

This paper is organized as follows: four possible color flux-tube structures for the heavy pentaquark states and the hamiltonian in the
color flux-tube model are given in Sec. II. The numerical calculations and discussions on the heavy petaquark states are presented in
Sec. III. A brief summary is given in the last section.

\section{color flux-tube structures and model hamiltonian}
Four possible color flux-tube structures of the pentaquark state $uudc\bar{c}$ are presented in FIG. 1, in which $q_i$ stands for
a light quark $u$ or $d$ and the codes of the quarks (antiquarks) $q$, $q$, $c$, $q$ and $\bar{c}$ are assumed to be 1, 2, 3, 4
and 5, respectively. Their positions are denoted as $\mathbf{r}_1$, $\mathbf{r}_2$, $\mathbf{r}_3$, $\mathbf{r}_4$, and $\mathbf{r}_5$,
$\mathbf{y}_i$ represents the $i$-th Y-shaped junction where three color flux-tubes meet. The color flux-tube structure (\textbf{1}) is,
a color singlet, a loose baryon-meson molecule state $[qqc]_{\mathbf{1}}[q\bar{c}]_{\mathbf{1}}$, the subscripts represent color dimensions.
The pentaquark states $P_c^+(4380)$ and $P_c^+(4450)$ were discussed in this picture due to their proximity to baryon-meson thresholds
within different theoretical framework~\cite{molecule}. The color flux-tube structures (\textbf{2}), (\textbf{3}) and (\textbf{4}) are
hidden color states. The pentaquark state with the flux-tubes structure (\textbf{2}) is a color octet state $[[qqc]_{\mathbf{8}}[q\bar{c}]_{\mathbf{8}}]_{\mathbf{1}}$, which generally has high energies due to a repulsive interaction between the
colored sub-clusters $[qqc]_{\mathbf{8}}$ and $[q\bar{c}]_{\mathbf{8}}$. The color flux-tube structure (\textbf{3}) is a diquark-diquark-antiquark state $[[qq]_{\bar{\mathbf{3}}}[cq]_{\bar{\mathbf{3}}}\bar{c}]_{\mathbf{1}}$, which interacts through the color force due to gluon exchange or flavor-dependent force due to meson exchange. The pioneer application of the diquark model applied to explain the structure of the pentaquark
$\Theta^+$ were done by Jaffe and Wilczek~\cite{jwdiquark}. The last structure is the so-called pentagonal state, which can be generated by
means of exciting two Y-shape junctions and a color flux-tube between $c$ and $q_1$ or $\bar{c}$ and $q_1$ from the vacuum based on the second
or third strucutre, respectively. One can suppose that the recombination of color flux-tubes is faster than the motion of the quarks because the
quarks in the constituent quark model are massive. Subsequently, the ends of five compound flux-tubes can meet each other in turn to establish a closed color flux-tube structure, a pentagon-$\mathbf{y}_1\mathbf{y}_3\mathbf{y}_2\mathbf{y}_4\mathbf{y}_5$. According to overall color singlet
and $SU(3)$ color coupling rule, the color flux-tube $\mathbf{y}_2\mathbf{y}_3$ is $\mathbf{8}$ dimension and others are $\mathbf{3}$ or $\bar{\mathbf{3}}$ dimension. It is worth mentioning that the counterpart of the pentagonal state in the QED world, the hydrocarbon $C_5H_5$ (or generally speaking $C_{2n+1}H_{2n+1}$, $n\in N$), does
not seem to exist.
\begin{figure}
\epsfxsize=3.6in \epsfbox{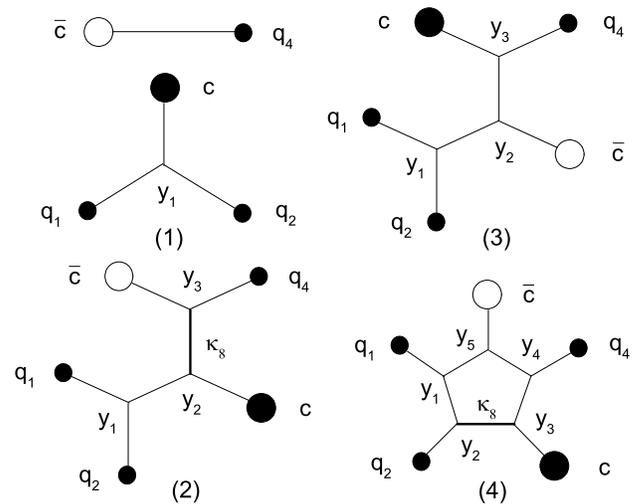}
\caption{Four possible color flux-tube structures for the pentaquark state $uudc\bar{c}$.}
\end{figure}

The interactions among quarks is one of the significant quantities for the study of the multiquark system in quark models. LQCD
investigations on mesons, baryons, tetraquark and pentaquark states reveal Y-shaped flux-tube structures~\cite{lattice}, which works
as a collective degree of freedom connecting all particles to form an overall color singlet hadron. The interactions obey the Coulomb
potential plus Y-type linear confinement potential proportional to the minimum of the sum of the lengthen of all color flux-tubes~\cite{lattice}.
A multiquark color flux-tube model has been developed based on the LQCD picture involving a multi-body confinement potential with a harmonic
interaction approximation, i.e., a sum of the square of the length of flux-tubes rather than a linear one is assumed to simplify the calculation~\cite{ping,ww}. The approximation is justified with the following two reasons: one is that the spatial variations in separation
of the quarks (lengths of the flux tube in different hadrons do not differ significantly, so the difference between the two functional
forms is small and can be absorbed in the adjustable parameter, the stiffness of color flux-tubes. The other is that we are using a
nonrelativistic dynamics in the study. As was shown long ago~\cite{goldman}, an interaction energy that varies linearly with separation
between fermions in a relativistic first order differential dynamics has a wide region in which a harmonic approximation is valid for the
second order (Feynman-Gell-Mann) reduction of the equations of motion. The comparative studies also indicated that the difference between
the quadratic confinement potential and the linear one is very small~\cite{ping,ww}.

Within the picture of color flux-tubes, the quadratic confinement potential is believed to be flavor independent~\cite{ground,b-bbar,flux}.
According to the color flux-tube structures of mesons and baryons in FIG. 1 (1), the confinement potential of mesons and baryons in the
color flux-tube model can be written as,
\begin{eqnarray}
V_{min}^C(2)&=&K(\mathbf{r}_4-\mathbf{r}_5)^2, \\
V^C(3)&=&K\left((\mathbf{r}_1-\mathbf{y}_1)^2+(\mathbf{r}_2-\mathbf{y}_1)^2+(\mathbf{r}_3-\mathbf{y}_1)^2\right).\nonumber
\end{eqnarray}
$K$ is the stiffness of the three-dimension color flux-tube. The minimum of the confinement potential of baryons can be obtained by taking
the variation of the confinement potential with respect to $\mathbf{y}_{1}$ and has therefore the following form,
\begin{eqnarray}
V_{min}^{C}(3)=
K\left(\left(\frac{\mathbf{r}_1-\mathbf{r}_2}{\sqrt{2}}\right)^2
+\left(\frac{2\mathbf{r}_3-\mathbf{r}_1-\mathbf{r}_2}{\sqrt{6}}\right)^2\right),
\end{eqnarray}
The confinement potential $V_i^{C}(5)$ ($i=1,2,3,4$) for the pentaquark states $uudc\bar{c}$ with the $i$-th color flux-tube structure
listed in FIG. 1 can be expressed as,
\begin{eqnarray}
V^{C}_1(5)&=&K\left( (\mathbf{r}_1-\mathbf{y}_1)^2
+(\mathbf{r}_2-\mathbf{y}_1)^2+(\mathbf{r}_{3}-\mathbf{y}_1)^2\right. \nonumber \\
&+&\left.(\mathbf{r}_4-\mathbf{r}_{5})^2\right),\\
V^{C}_2(5)&=&K\left( (\mathbf{r}_1-\mathbf{y}_{1})^2
+(\mathbf{r}_2-\mathbf{y}_{1})^2+(\mathbf{r}_{3}-\mathbf{y}_{2})^2\right. \nonumber \\
&+&\left.(\mathbf{r}_4-\mathbf{y}_{3})^2+(\mathbf{r}_5-\mathbf{y}_{3})^2+(\mathbf{y}_{1}-\mathbf{y}_{2})^2\right.\nonumber\\
&+&\left.\kappa_8(\mathbf{y}_{2}-\mathbf{y}_{3})^2\right),
\end{eqnarray}
\begin{eqnarray}
V^{C}_3(5)&=&K\left( (\mathbf{r}_1-\mathbf{y}_{1})^2
+(\mathbf{r}_2-\mathbf{y}_{1})^2+(\mathbf{r}_{3}-\mathbf{y}_{2})^2\right. \nonumber \\
&+&\left.(\mathbf{r}_4-\mathbf{y}_{3})^2+(\mathbf{r}_5-\mathbf{y}_{3})^2+(\mathbf{y}_{1}-\mathbf{y}_{2})^2\right.\nonumber\\
&+&\left.(\mathbf{y}_{2}-\mathbf{y}_{3})^2\right),\\
V^{C}_4(5)&=&K\sum_{i=1}^5(\mathbf{r}_i-\mathbf{y}_{i})^2
+K\left ((\mathbf{y}_1-\mathbf{y}_{2})^2+(\mathbf{y}_{2}-\mathbf{y}_{3})^2\right. \nonumber \\
&+&\left.(\mathbf{r}_3-\mathbf{y}_{4})^2+\kappa_8(\mathbf{y}_4-\mathbf{y}_{5})^2+(\mathbf{y}_5-\mathbf{y}_1)^2\right).\nonumber
\end{eqnarray}
The relative stiffness parameter $\kappa_8$ of the color $\mathbf{8}$ dimension flux-tube is $\kappa_8=\frac{C_8}{C_3}$~\cite{kappa},
where $C_8$ and $C_3$ are the eigenvalues of the Casimir operator associated with the $SU(3)$ color representation on either end of
the color flux tube, namely $C_3=\frac{4}{3}$ and $C_8=3$.

The confinement potential $V^{C}_i(5)$ can be simplified into the sum of five independent harmonic oscillators by taking the variation
with respect to $\mathbf{y}_i$ and then diagonalizing the matrix of the confinement potential. Finally, the confinement potential can
be expressed as
\begin{eqnarray}
V^{C}_i(5)=K\sum_{j=1}^{5}k_{ij}\mathbf{R}^2_{ij},
\end{eqnarray}
For the sake of simplicity, the eigenvalue $k_{ij}$ can be written in the form
\begin{eqnarray}
k=\left (
\begin{array}{cccccc}
 1      &   1      &    2      &   0     &  0  \\
 0.406  &   1      &    0.820  &   1     &  0  \\
 1      &   0.333  &    0.714  &   1     &  0  \\
 0.580  &   0.783  &    0.638  &   0.862 &  0  \\
\end{array}
\right ),
\end{eqnarray}
$\mathbf{R}_{ij}$ is an eigenvector corresponding to the eigenvalue $k_{ij}$. A vector $\mathbf{R}_i$ for the $i$-th color flux-tube
structure in FIG. 1 can be constructed as $\mathbf{R}_i=(\mathbf{R}_{i1}~\mathbf{R}_{i2}~\mathbf{R}_{i3}~\mathbf{R}_{i4}~\mathbf{R}_{i5})^T$
and $\mathbf{R}_{i}=\mathbf{M}_i\mathbf{r}$, the vector $\mathbf{r}=(\mathbf{r}_1~\mathbf{r}_2~\mathbf{r}_3~\mathbf{r}_4~\mathbf{r}_5)^T$.
The $i$-th transformation matrix $\mathbf{M}_i$ has the following forms,
\begin{eqnarray}
\mathbf{M}_1&=&\left (
\begin{array}{cccccc}
 0.707 &  -0.707  &  0     &   0     &   0      \\
 0.408 &   0.408  & -0.816 &   0     &   0      \\
 0     &   0      &  0     &   0.707 &  -0.707  \\
 0.365 &   0.365  &  0.365 &  -0.548 &  -0.548  \\
 0.447 &   0.447  &  0.447 &   0.447 &   0.447  \\
\end{array}
\right ),  \nonumber \\
\mathbf{M}_2&=&\left (
\begin{array}{cccccc}
 0.537 &   0.537  & -0.198 &  -0.438  & -0.438  \\
 0.707 &  -0.707  &  0     &   0      &  0      \\
 0.107 &   0.107  & -0.872 &   0.329  &  0.329  \\
 0     &   0      &  0     &  -0.707  &  0.707  \\
 0.447 &   0.447  &  0.447 &   0.447  &  0.447  \\
\end{array}
\right ), \nonumber \\
\mathbf{M}_3&=&\left (
\begin{array}{cccccc}
 0.707 &  -0.707  &  0     &   0      &  0      \\
 0.5   &   0.5    &  0     &  -0.5    & -0.5    \\
 0.224 &   0.224  & -0.894 &   0.224  &  0.224  \\
 0     &   0      &  0     &  -0.707  &  0.707  \\
 0.447 &   0.447  &  0.447 &   0.447  &  0.447  \\
\end{array}
\right ),\nonumber \\
\mathbf{M}_4&=&\left (
\begin{array}{cccccc}
 0.632 &   0.195  & -0.512 &  -0.512  &  0.195  \\
 0.632 &  -0.512  &  0.195 &   0.195  & -0.512  \\
 0     &   0.688  &  0.162 &  -0.162  & -0.688  \\
 0     &   0.162  & -0.688 &   0.688  & -0.162  \\
 0.447 &   0.447  &  0.447 &   0.447  &  0.447  \\
\end{array}
\right ).\nonumber
\end{eqnarray}

One-gluon-exchange interaction (coulomb interaction plus color-magnetic interaction) is very important because of the responsibility
for the mass splitting in the hadron spectra, it takes the standard form and can be read as~\cite{vijande}
\begin{eqnarray}
V_{ij}^{G} & = & {\frac{\alpha_{s}}{4}}\mathbf{\lambda}_{i}\cdot\mathbf{\lambda}_{j}\left({\frac{1}{r_{ij}}}-
{\frac{2\pi\delta(\mathbf{r}_{ij})\mathbf{\sigma}_{i}\cdot\mathbf{\sigma}_{j}}{3m_im_j}}\right),
\end{eqnarray}
where $m_i$ is the mass of the $i$-th quark (antiquark), the symbols $\mathbf{\lambda}$ and $\mathbf{\sigma}$ are the color SU(3)
Gell-man and spin $SU(2)$ Pauli matrices, respectively. The running strong coupling constant $\alpha_s$ takes the following form
\begin{equation}
\alpha_s(\mu_{ij})=\frac{\alpha_0}{\ln\frac{\mu_{ij}^{2}}{\Lambda_0^2}},
\end{equation}
The function $\delta(\mathbf{r}_{ij})$ should be regularized; the regularization is justified based on the finite size of the constituent
quark and should, therefore, be flavor dependent~\cite{weistein},
\begin{equation}
\delta(\mathbf{r}_{ij})=\frac{1}{4\pi r_{ij}r_0^2(\mu_{ij})}e^{-r_{ij}/r_0(\mu_{ij})},
\end{equation}
where $\mu_{ij}$ is the reduced mass of two interacting particles $q_i$ (or $\bar{q}_i$) and $q_j$ (or $\bar{q}_j$),
$r_0(\mu_{ij})=r_0/\mu_{ij}$.

To sum up, the color flux-tube model Hamiltonian $H_n$ for mesons, baryons and pentaquark states can be universely expressed as,
\begin{eqnarray}
H_n = \sum_{i=1}^n \left(m_i+\frac{\mathbf{p}_i^2}{2m_i}
\right)-T_c+\sum_{i>j}^n V^{G}_{ij}+V^{C}_{min}(n),
\end{eqnarray}
$T_c$ is the center-of-mass kinetic energy of the state, $\mathbf{p}_i$ is the momentum of the $i$-th quark (antiquark), respectively.
The tensor and spin-orbit forces between quarks are omitted in the present calculation because, for the lowest energy states which we
are interested in here, their contributions are small or zero.

\section{numerical calculations and discussions}

The stiffness $K$ of the three-dimension color flux tube is considered as a fixed parameter and taken to be 700 MeV fm$^{-2}$. The seven
adjustable model parameters, $m_u$, $m_s$, $m_c$, $m_b$, $\Lambda_0$, $r_0$, $\alpha_0$, and their errors can be fixed by fitting the mass
spectra of ground states of heavy mesons and baryons using Minuit program, which are presented in Table I and Table II, respectively.
The mass spectra can be obtained by solving the two-body and three-body Schr\"{o}dinger equation
\begin{eqnarray}
(H_n-E_{n})\Phi_{IJ}^{n}=0.
\end{eqnarray}
with Rayleigh-Ritz variational principle, where $n=2$ and $3$, the details of the construction of the wave functions of baryons and mesons
can be found in the papers~\cite{ground,b-bbar}. The mass errors of heavy mesons and baryons $\Delta E_n$ introduced by the parameter uncertainty $\Delta x_i$ can be calculated by the formula of error propagation,
\begin{eqnarray}
\Delta H_n &= & \sum_{i=1}^{7}\left|{\frac{\partial{H_n}}{\partial{x_i}}}\right|\Delta x_i, \\
\Delta E_n & \approx & \left <\Phi_{IJ}^{n}\left|\Delta
H_n\right|\Phi_{IJ}^{n}\right >.
\end{eqnarray}
where $x_i$ and $\Delta x_i$ represent the $i$-th adjustable parameter and it's error, respectively, which are listed in Table II.

\begin{table}[ht]
\caption{Adjustable parameters in the color flux-tube model.
(units: $m_u$, $m_s$, $m_c$, $m_b$, $\Lambda_0$, MeV; $r_0$, MeV$\cdot$fm; $\alpha_0$, dimensionless)\label{para}}
\begin{tabular}{cccccc}
\hline\hline
Parameters     & ~~~~$x_i$~~~~    &  ~~$\Delta x_i$~~  &    Parameters   &   ~~~~$x_i$~~~~   &  ~$\Delta x_i$~ \\
$m_{u}$        &      230.06      &      0.28530       &    $\alpha_0$   &      4.6945       &      0.00499    \\
$m_{s}$        &      473.29      &      0.23195       &    $\Lambda_0$  &      30.241       &      0.03927    \\
$m_{c}$        &      1701.3      &      0.30672       &    $r_0$        &      81.481       &      0.05267    \\
$m_{b}$        &      5047.0      &      0.44204       &                                                       \\
\hline\hline
\end{tabular}
\caption{Ground state heavy-meson and baryon spectra, unit in MeV.\label{meson}}
\begin{tabular}{ccccccccccccc}
\hline\hline
~~States~~     & $E_2\pm\Delta E_2$ &   ~~PDG~~    &     ~~States~~     &  $E_2\pm\Delta E_2$   &    ~~PDG~~  \\
$D^{\pm}$      &     1879$\pm$2     &     1869     &     $D^*$          &     2039$\pm$2        &     2007    \\
$D_s^{\pm}$    &     1952$\pm$2     &     1968     &     $D_s^*$        &     2144$\pm$2        &     2112    \\
$\eta_c$       &     2949$\pm$3     &     2980     &     $J/\Psi$       &     3168$\pm$2        &     3097    \\
$B^0$          &     5285$\pm$2     &     5280     &     $B^*$          &     5343$\pm$2        &     5325    \\
$B_s^0$        &     5352$\pm$2     &     5366     &     $B_s^*$        &     5429$\pm$2        &     5416    \\
$B_c$          &     6254$\pm$2     &     6277     &     $B_c^*$        &     6396$\pm$2        &     ...     \\
$\eta_b$       &     9374$\pm$3     &     9391     &     $\Upsilon(1S)$ &     9536$\pm$3        &     9460    \\
\hline
~~States~~     & $E_3\pm\Delta E_3$ &   ~~PDG~~    &     ~~States~~     &  $E_3\pm\Delta E_3$   &    ~~PDG~~  \\
$N$            &     $945\pm4$      &     939      &     $\Lambda$      &    $1128\pm4$         &     1115    \\
$\Sigma$       &     $1204\pm3$     &     1195     &     $\Xi$          &    $1345\pm3$         &     1315    \\
$\Delta$       &     $1230\pm3$     &     1232     &     $\Sigma^*$     &    $1391\pm3$         &     1385    \\
$\Xi^*$        &     $1537\pm2$     &     1530     &     $\Omega$       &    $1677\pm2$         &     1672    \\
$\Lambda_c^+$  &     $2278\pm4$     &     2285     &     $\Sigma_c$     &    $2437\pm3$         &     2445    \\
$\Sigma_c^*$   &     $2508\pm3$     &     2520     &     $\Xi_c$        &    $2460\pm3$         &     2466    \\
$\Xi_c^*$      &     $2626\pm2$     &     2645     &     $\Omega_c^0$   &    $2703\pm2$         &     2695    \\
$\Omega_c^{0*}$&     $2774\pm2$     &     2766     &     $\Lambda_b^0$  &    $5596\pm4$         &     5620    \\
$\Sigma_b$     &     $5786\pm3$     &     5808     &     $\Sigma_b^*$   &    $5812\pm3$         &     5830    \\
$\Xi_b$        &     $5765\pm3$     &     5790     &     $\Xi_b^*$      &    $5917\pm3$         &     ...     \\
$\Omega_b^-$   &     $6034\pm2$     &     6071     \\
\hline\hline
\end{tabular}
\end{table}

Next, let's discuss the pentaquark states $uudc\bar{c}$ within the framework of diquark-diquark-antiquark $[ud][cd]\bar{c}$.
The diquarks $[ud]$ and $[cu]$ are considered as no internal orbital excitation, and the angular excitation $L$ is assumed to
occur only between two subclusters $[ud\bar{c}]$ and $[cd]$ if orbital excitation is permitted, which induces the lower relative
kinetic energy between the two subclusters because of the bigger reduced mass. Therefore, the parity of the pentaquark states
$uudc\bar{c}$ is $(-1)^{L+1}$. In this way, the wave function of the pentaquark states $uudc\bar{c}$ with quantum numbers $IJ^P$
can be expressed as,
\begin{eqnarray}
\Phi^{5}_{IJ}=\sum_zc_z\left[\mathcal{A}_{uud}\left[\psi_{c_1s_1f_1}^{[ud]}\psi_{c_2s_2f_2}^{[cu]}\psi_{c_3s_3f_3}^{\bar{c}}
\right]_{IS}\psi^G_{LM}\right]_{IJ},
\end{eqnarray}
The intermediate quantum numbers $c_i$, $s_i$ and $i_i$ stand for the color, spin and isospin, respectively, the subscript $i=1,2$
and $3$. The details of the wave functions $\psi_{c_i, s_i, i_i}$ are omitted here. All [~]'s represents all possible Clebsch-Gordan (C-G)
coupling. The $c_z$ is a C-G coefficient, $z=\{c_1, s_1, i_1,c_2, s_2, i_2,c_3, s_3, i_3, I, S, J\}$.

The $\psi^G_{LM}$ is the total spatial wave function of the pentaquark states, in which the part of the identical particles $uud$
are assumed to be symmetrical because we are interested in the low energy states here. In this way, the color-spin-isospin wave
functions of the three identical quarks $uud$ should be antisymmetrical due to Pauli principle, anti-symmetrized operator $\mathcal{A}_{uud}=1-P_{14}-P_{24}$, which only operates on color, spin and isospin parts of the wave function because the
orbital part is symmetrical.

In order to obtain the symmetrical spatial wave functions of three identical quarks $uud$, we can define a set of cyclic Jacobi
coordinates $\mathbf{r}_{ij}$, $\mathbf{R}_k$, $\mathbf{T}_{ij}$ and $\mathbf{Q}_{ijk}$ for the cyclic permutations of $(i,j,k)=(1,2,4)$,
\begin{eqnarray}
\mathbf{r}_{ij}&=&\mathbf{r}_i-\mathbf{r}_j,~\mathbf{R}_k~=~\mathbf{r}_3-\mathbf{r}_k, ~
\mathbf{T}_{ij}~=~\frac{\mathbf{r}_i+\mathbf{r}_j}{2}-\mathbf{r}_5, \nonumber\\
\mathbf{Q}_{ijk}&=&\frac{m_{u}\mathbf{r}_i+m_u\mathbf{r}_j+m_c\mathbf{r}_5}{2m_u+m_c}-\frac{m_c\mathbf{r}_3+m_u\mathbf{r}_k}{m_u+m_c}.
\end{eqnarray}
In this way, the total orbital wave function $\psi^G_{LM}$ can be expressed as
\begin{eqnarray}
\psi^G_{LM}=\sum_{i,j,k}\phi^G_{00}(\mathbf{r}_{ij})\phi^G_{00}(\mathbf{R}_k)\phi^G_{00}(\mathbf{T}_{ij})\phi^G_{LM}(\mathbf{Q}_{ijk}),
\end{eqnarray}
The Gaussian expansion method (GEM) has been proven to be rather powerful in solving a few-body problem~\cite{gem}, in which the relative
motion wave function $\phi^G_{lm}(\mathbf{x})$ can be expanded as the superposition of many single Gaussian functions with different
size $\nu_k$
\begin{eqnarray}
\phi^G_{lm}(\mathbf{x})=\sum_{k=1}^{k_{max}}c_{k}N_{kl}x^{l}e^{-\nu_{k}x^2}Y_{lm}(\hat{\mathbf{x}})
\end{eqnarray}
The expansion coefficient $c_k$ can be determined by the dynamics of the pentaquark system. The other details of the wavefunction $\phi^G_{lm}(\mathbf{x})$ can be found in the paper~\cite{gem}.

The color flux-tube structure specifies how the colors of quarks and antiquarks are coupled to form an overall color singlet. Similarly,
the color wave functions of the baryon-meson molecules and color octet states can be constructed in the model study. It is, however,
difficult to construct the color wave function of the novel color flux-tube structure, pentagonal state, only using quark degrees of
freedom if no explicit gluon is introduced in the quark models. In fact, it is difficult to introduce an explicit gluon degree of freedom
in the nonrelativistic quark models because of the zero mass of gluons. Furthermore, the predictive power of quark models will be reduced
due to the increase of model parameters even if the constituent gluons can be introduced. The wave function of the pentagonal structure is
therefore assumed to be the same as that of the diquark-diquark-antiquark structure to estimate the energy of the pentaquark states with
pentagonal structure in the present work.

Subsequently, the color flux-tube model with the model parameters listed in the Table II is extended to study the properties
of the heavy pentaquark states. The converged numerical results $E_5$'s can be obtained by solving a five-body Schr\"{o}dinger
equation
\begin{eqnarray}
(H_5-E_{5})\Phi_{IJ}^{5}=0.
\end{eqnarray}
with Rayleigh-Ritz variational principle under the conditions of $k_{max}=5$, $r_1=0.3$ fm and $r_{k_{max}}=2.0$ fm.
The error $\Delta E_5$ can be calculated as $\Delta E_2$ and $\Delta E_3$.

\begin{table}
\caption{The energies $E_5\pm\Delta E_5$ of the ground states of the heavy pentaquark states $uudc\bar{c}$, $uudb\bar{c}$,
$uudc\bar{b}$ and $uudb\bar{b}$ with $J^P$ and three color structures in the color flux-tube model, unit in MeV.}
\begin{tabular}{cccccccccc} \hline\hline
     Flavors    &       ~$J^P$~       &   ~~~Octet~~~   & ~~Diquark~~   &  ~Pentagon~  &    Candidate   \\
                &   $\frac{1}{2}^-$   &   $4402\pm5$    &  $4344\pm5$   &  $4303\pm5$  &      ...       \\
 $uudc\bar{c}$  &   $\frac{3}{2}^-$   &   $4473\pm5$    &  $4405\pm5$   &  $4369\pm5$  &  $P_c^+(4380)$ \\
                &   $\frac{5}{2}^-$   &   $4616\pm4$    &  $4567\pm4$   &  $4516\pm4$  &  $P_c^+(4450)?$ \\

                &   $\frac{1}{2}^-$   &   $7612\pm5$    &  $7609\pm5$   &  $7564\pm5$  &      ...       \\
  $uudb\bar{c}$ &   $\frac{3}{2}^-$   &   $7634\pm5$    &  $7631\pm5$   &  $7587\pm5$  &      ...       \\
                &   $\frac{5}{2}^-$   &   $7812\pm4$    &  $7788\pm4$   &  $7738\pm4$  &      ...       \\

                &   $\frac{1}{2}^-$   &   $7650\pm5$    &  $7618\pm5$   &  $7573\pm5$  &      ...       \\
  $uudc\bar{b}$ &   $\frac{3}{2}^-$   &   $7702\pm5$    &  $7658\pm5$   &  $7613\pm5$  &      ...       \\
                &   $\frac{5}{2}^-$   &   $7817\pm4$    &  $7790\pm4$   &  $7740\pm4$  &      ...       \\

                &   $\frac{1}{2}^-$   &   $10747\pm5$   &  $10616\pm6$  &  $10587\pm6$ &      ...       \\
  $uudb\bar{b}$ &   $\frac{3}{2}^-$   &   $10767\pm5$   &  $10622\pm5$  &  $10592\pm5$ &      ...       \\
                &   $\frac{5}{2}^-$   &   $10947\pm5$   &  $10935\pm5$  &  $10892\pm5$ &      ...       \\
\hline\hline
\end{tabular}
\end{table}

The energies $E_5\pm\Delta E_5$ of the ground states of the heavy pentaquark states $uudc\bar{c}$, $uudc\bar{b}$, $uudb\bar{c}$ and
$uudb\bar{b}$ with three different color flux-tube structures, diquark-diquark-antiquark (Diquark), color octet state (Octet) and
pentagonal state (Pentagon), under the assumptions of total spin $S=\frac{1}{2}$, $S=\frac{3}{2}$ and $S=\frac{5}{2}$ are systematically
calculated and presented in Table III. Their corresponding $J^P$ are therefore $\frac{1}{2}^-$, $\frac{3}{2}^-$ and $\frac{5}{2}^-$
because of $L=0$. It can be seen from Table III that the energy errors $\Delta E_5$ are very small, just several MeVs. The bigger
the angular momentum $J$, the higher the energy $E_5$ of the pentaquark states with the same quark content. The energies of the
pentaquark states with the same flavor and quantum numbers but three different color flux-tube structure are close, the difference
among them mainly comes from the contribution of one-gluon-exchange interaction. The previous investigation on the six-quark state
indicated that the energy difference among different color flux-tube structures is very small if one-gluon-exchange interaction is not involved~\cite{ping}. These different color flux-tube structures with the same flavor can therefore be called QCD isomers analogous
to the isomers in QED world, which have different chemical bond structure but same atom constituent. The energy of the pentaquark
states with a ring-like color flux-tube structure is lower than that of the state with chain-like structures in the color flux-tube
model with quadratic confinement potential because the ring-like structure is easier to shrink into a compact multiquark state than
chain-like structures. In this way, the energy of the pentaquark with the pentagonal structure is lower than that of the diquark
structure. However, the energy of color octet state is higher than the diquark structure mainly because of a repulsive one-gluon-exchange
interaction in the two colored octet sub-clusters. In addition, it is worth mentioning that baryon-meson molecule configuration can
not be formed in the color flux-tube model because there does not exist binding mechanism except one-gluon-exchange interaction, which
is not enough to bind a baryon and a meson into a loose hadron molecule state.

The energy of the state $uudc\bar{c}$ with the pentagonal structure and $J^P=\frac{3}{2}^-$ is $4369\pm5$ MeV in the color flux-tube model,
see Table III, which is highly consistent with experimental data of the state $P^+_c(4380)$. It is therefore possible to explain the state
$P^+_c(4380)$ as the state $uudc\bar{c}$ with $J^P=\frac{3}{2}^-$, which is supported by a large number of theoretical studies~\cite{molecule,
diquark,diquark-triquark}. The energy of the state $uudc\bar{c}$ with the pentagonal structure and $J^P=\frac{5}{2}^-$ is $4516\pm4$ MeV in
the color flux-tube model, which is a litter higher than that of the state $P^+_c(4450)$ and, however, agrees with the conclusions in several researches~\cite{positive1,positive2,positive3}. The energies of the states $uudc\bar{c}$ with positive parity $(L=1)$ and total spin $S=\frac{1}{2}$, $\frac{3}{2}$ and $\frac{5}{2}$ are, respectively, $4602\pm5$ MeV, $4632\pm5$ MeV and $4781\pm4$ MeV in the color flux-tube model (the spin-orbit interaction is very weak and therefore not taken into account here~\cite{zc}), which are much higher than the energies of the two $P_c$ states
and close to the prediction on the states in the work~\cite{positive2}. Therefore, these positive parity states should not be the main component of the two $P^+_c$ states in the color flux-tube model. In this way, the optimum assignment of the main component of the state $P^+_c(4450)$ from the mass seems to be the state $uudc\bar{c}$ with $J^P=\frac{5}{2}^-$ in the color flux-tube model. However, the negative parity is contradictive with 
the assignment of the opposite parity of the two $P^+_c$ states reported by LHCb Collaboration. The state $P^+_c(4450)$ is therefore difficult to be accommodated into the color flux-tube model and worth of further research in the future.

The expected lowest energy of the state $uudc\bar{c}$ with the pentagonal structure and $J^P=\frac{1}{2}^-$ is $4303\pm5$ MeV in the color
flux-tube model, which is close to the prediction on the state in the work~\cite{positive2}. The energies of the hidden beauty pentaqquark
states $uudb\bar{b}$ with different quantum numbers and structures are similarly estimated in the color flux-tube model, which are lower than
those of the states in the researches~\cite{positive3,vkip}. The energies of the states $uudb\bar{b}$ in the color flux-tube model should be underestimated mainly because of the strong Coulomb attractive interaction due to the small distance among heavy quarks, the details can be
found in our previous work~\cite{ground}. In addition, the pentaquark states $uudc\bar{b}$ and $uudb\bar{c}$ are also predicted in the color
flux-tube model. The pentaquark states $uudc\bar{b}$ and $uudb\bar{c}$ with $J^P=\frac{5}{2}^-$ almost share the same energies. For $ J^P=\frac{1}{2}^-$ and $J^P=\frac{3}{2}^-$, the energies of the states $uudc\bar{b}$ is a little higher than those of the states $uudb\bar{c}$.

The main component analysis of the two $P^+_c$ states is only based on the mass calculation. The crucial test of the main components should
be determined by the systematic study of their decays, which involves a channel coupling calculation containing all possible color flux-tube
structures and is left for the further research in the future. The five-body color flux-tube is a collective degree of freedom, which
acts as a dynamical mechanism and plays an important role in the formation and decay of those compact pentaquark states. Different
topological structures of color flux tubes induce the diversity of inner color configurations in the pentaquark states. In general,
the pentaquark states should be the mixtures of all possible color flux-tube structures, especially within the range of confinement
(about 1 fm). These different structures can transform one another, which can be understood here that the gluon field readjusts immediately
to its minimal configuration. In this way, the flip-flop of color flux-tube structures may induce a color structure resonance, which
can be called a color confined, multiquark resonance state~\cite{resonance}.

\section{summary}

Within the framework of the color flux-tube model including a five-body confinement potential, a novel color flux-tube structure,
pentagonal structure, for pentaquark states is presented because of the observation of the two $P^+_c$ states. The pentagonal structure
provide a new insight into the inner structure of the pentaquark states. Numerical calculations on the heavy pentaquark states indicate
that three color flux-tube structures, diquark, octet and pentagonal states, have the close masses. The pentagonal structure has lowest
energy because the ring-like structure is easier to shrink into a compact multiquark state than chain-like structure. These different
color flux-tube structures with the same flavor can be called QCD isomers analogous to QED isomers. The five-body confinement potential
basing on the color flux-tube as a collective degree of freedom plays an important role in the formation of those compact heavy pentaquark
states.

The main component of the state $P^+_c(4380)$ can be described as a compact pentaquark state $uudc\bar{c}$ with the pentagonal structure
and $J^P=\frac{3}{2}^-$ in the color flux-tube model. Although the lowest mass of the state $uudc\bar{c}$ with $J^P=\frac{5}{2}^-$ is not
far from the experimental data of the state $P^+_c(4450)$, it should not be a good candidate of the main component of the state $P^+_c(4450)$
because of the same parity with the state $P^+_c(4380)$ in the color flux-tube model.  The states $uudc\bar{c}$ with positive parity have
masses much higher than those of the states $P^+_c(4380)$ and $P^+_c(4450)$ in the color flux-tube model. It is therefore hard to describe
the state $P^+_c(4450)$ in the color flux-tube model. The heavy pentaquark states $uudc\bar{b}$, $uudb\bar{c}$ and $uudb\bar{b}$ are predicted
in the color flux-tube model. These mass calculations on the heavy pentaquark states may be useful for planning future experiments and studying manifestly exotic baryon states to complete the picture of exotic baryons.

\acknowledgments
{This research is partly supported by the NSFC under contracts Nos. 11305274, 11175088, 11035006, 11205091, and the Chongqing Natural Science Foundation under Project No. cstc2013jcyjA00014.}

\end{document}